\documentclass[aip]{revtex4-1}

\usepackage[utf8]{inputenc}
\usepackage{amsmath}
\usepackage{amssymb}
\usepackage{graphicx}
\usepackage{color,xcolor}
\usepackage{hyperref} 
\usepackage{wrapfig}
\usepackage{multirow}
\usepackage{caption}
\usepackage{subcaption}
\usepackage{booktabs}

\usepackage{comment} 
\usepackage{epstopdf}
\usepackage{float}
\usepackage[font=normal]{subcaption}

\usepackage[top=0.9in, left=0.9in,bottom=1.2in,right=0.9in]{geometry}

\usepackage{cleveref}

\usepackage[sort&compress]{natbib}

\newcommand{\xx}{\mathbf{x}}
\newcommand{\KK}{\mathbf{K}}
\newcommand{\bfm}{\mathbf{M}}

\newcommand{\bveps}{\boldsymbol{\varepsilon}}

\newcommand{\rmd}{\mathrm{d}}

\graphicspath{{./Images}}

\begin{document}

\title{On the low frequency flexural band gaps of a metamaterial plate with low porosity} 



\author{Chaitanya Morey}
 \email{me20d017@smail.iitm.ac.in}
 \affiliation{Department of Mechanical Engineering, Indian Institute of Technology Madras, Chennai-600036.}
\author{Sundararajan Natarajan}%
 \email{snatarajan@iitm.ac.in.}
\affiliation{Department of Mechanical Engineering, Indian Institute of Technology Madras, Chennai-600036.
}%

\author{Chandramouli Padmanabhan}
 \email{mouli@iitm.ac.in}
\affiliation{%
Department of Mechanical Engineering, Indian Institute of Technology Madras, Chennai-600036.
}%

\date{\today}

\begin{abstract}
This paper demonstrates numerically and experimentally that it is possible to tailor flexural band gaps in the low-frequency regime by appropriate choice of cutout characteristics. The finite element method is used to obtain the numerical dispersion relation and band gaps. The influence of the cutout's shape, size, and location on the band gap is systematically studied. The study demonstrates that the cutout should pass through the center of the unit cell, and a large aspect ratio is required to introduce flexural band gaps in the low-frequency regime. This is validated by experiments on a finite plate with 3 $\times$ 3 unit cells. 
\end{abstract}

\maketitle

\section{Introduction}
Metamaterials are engineered periodic structures with tailored features that has properties not commonly found in conventional materials, such as negative mass density, negative Poisson's ratio, high specific stiffness and egative thermal expansions.~\cite{ai2017metamaterials,tan2019novel,gao2020ultrawide}. The properties of the metamtaterials are derived from the geometric design, which could be tailored according to the needs. By intelligently engineering the macro- and/or the micro-structure, the metamaterials can be used in vibration isolation \cite{yan2022propagation}, noise filtering\cite{mir2023metamaterials}, wave guiding \cite{li20214d}, seismic isolation \cite{jia2010new}, to name a few. This is due to the presence of `band gaps', which attenuates the wave passing through the structure. The band gaps could be of either electromagnetic waves, elastic waves or acoustic waves. Thanks to the advancements in additive manufacturing, realization of band gaps is now possible. 

Xiao et al,~\cite{xiaozeng2008} studied the flexural vibration in thin plates with two-dimensional ternary locally resonant structures. Xiong et al., ~\cite{xiongxu2023} designed periodic elastic metamaterial by the locally resonant band-gap mechanism. An asymptotic optimization method was employed to achieve required band gaps. Li et al.,~\cite{lidou2019} demonstrated that the presence of low-frequency band gap in thick elastic steel plate. This was achieved by adopting double-sided composite stepped resonators, which contain an array of rubber fillers embedded in a thick steel plate. Zhang et al.,~\cite{zhangxu2024} proposed a tunable pneumatic metamaterial plate with airbag local resonators to widen the band gap below 100 Hz. Xue et al.,~\cite{xuezhang2024} proposed a plate-type local resonator integrated with a base plate for vibration location and suppression. Yin et al.,~\cite{yinzhu2024} designed a new three-dimensional elastic metamaterial with surface resonant units using 3D printing technology. Chen et al.,~\cite{chenzi2021} employed a sandwich construction for isolating low-frequency vibration and flexural wave propagation in ship structures. Dal Paggetto and Serpa~\cite{dalpaggettoserpa2021} studied band gaps in a ternary periodic metamaterial using a plane wave expansion method. The plate kinematics was based on Mindlin plate theory, and the unit cell had the following materials: acrylonitrile-butadiene-styrene (ABS), low-density polyethylene, and lead. Li et al.,~\cite{lizhang2023} proposed a multi-band gap metamaterial for vibration suppression by integrating membrane-mass structures into a honeycomb sandwich structure.

In recent years, the investigation of porous metamaterials has increased due to its lightweight and ease of manufacturing. The effects of porous fractal phononic crystals and hierarchy on band gaps have also been studied. Wang et al.~\cite{wang2016band} examined the impact of fractal hierarchy on the in-plane wave band structure of Sierpinski triangular fractal porous phononic crystal (FPPC). Two types of Sierpinski triangle fractals were considered in their study: an equilateral triangle and a right-angled isosceles triangle. For the same level of fractal hierarchy, the Sierpinski equilateral triangle FPPC has wider band gaps. At the same time, at lower porosity (15\%), the first band opens, and also, a relatively lower band gap can be obtained with Sierpinski right-angle isosceles FPPC with a porosity of 70\%. The lowest band achieved is around 27 kHz for a unit cell with a dimension of 20 mm; the material used is aluminum. But this is achieved only at a high porosity of 60\% or more. Liu et al. \cite{liu2013influence} investigated the influence of T-square fractal holes on the in-plane band structure. The T-square fractal unit cell at the second level of the hierarchy has the same geometry used by Li et al.~\cite{li2023broadband}. At the fourth level, where the porosity is almost 80\%, the lowest band gap achieved is around the 5$^{\rm th}$/6$^{\rm th}$ mode. For lower porosities, the band gaps appear at higher modes. The unit cell is 10 mm in dimension and aluminum is used for in-plane studies. Shi et al.~\cite{shi2019elastic} studied the structural hierarchy of in-plane and out-of-plane band structures with tetrad elliptical patterns in perforated plates. They obtained band gaps for three levels of perforations; level 2 showed the maximum number of band gaps for a 52.3\% porosity. It must be noted that the lowest band gap occurs between the 10/11$^{\rm th}$ modes and the numerical results were validated experimentally. 

Jin et al.~\cite{jin2022deep}  studied the flexural band gap properties of corrugated plates sandwiched between two plates with spiral holes. While they demonstrated low-frequency band gaps, the structure developed is quite complex, with a combination of aluminum and resin material in the spiral pockets. Das et al.~\cite{das2021vibration} investigated the flexural band gaps of plates with different periodic cavity shapes  (circular, square, vertical rectangle and horizontal rectangle). The simulation results were validated with experiments. It was observed that changing the aspect ratio significantly affects bands; by decreasing the aspect ratio, the width of the band gap can be increased. Around 750 Hz, the lowest band was obtained with a horizontal rectangle-shaped cavity. Javid et al. ~\cite{javid2016architected} are possibly among the earliest researchers to look at the effects of low porosity levels on band gaps of periodic structures. They examined alternating orthogonal elliptical holes with 1\%, 5\%, and 10\% porosity levels. High aspect ratios enable them to demonstrate band gaps beyond the 6$^{\rm th}$ mode for 5\% porosity.  Tian et al. \cite{tian2020perforation} studied the orientation effect on the band gaps of the low-porosity perforated plates. The numerical results were verified with experiments, and the lowest in-plane band gap obtained is in the 3.4--4.1 kHz frequency range for 4.8\% porosity. Chen et al.~\cite{chen2018low} proposed a new auxetic perforated metamaterial with low porosity that shows negative Poisson’s ratio and band gap properties. The results show that the proposed configuration has enhanced tensile strength, and the chord length of the perforation has a significant impact on widening the band gap characteristics. 

From the discussions above, it is clear that one can generate wider band gaps either by higher porosity structures or by using binary or ternary material configurations. Moreover, the band gaps were reported beyond the 7$^{\rm th}$ or 8$^{\rm th}$ mode of a unit cell, primarily focusing on in-plane wave propagation. Only a few papers have been published focusing on periodic structures with low porosity. Again, in these papers, it is found that band gaps start appearing only after the 6$^{\rm th}$ or the 7$^{\rm th}$ mode. The challenge remains as to how to get these band gaps to lower modes, for example, in the 2$^{\rm nd}$ or 3$^{\rm rd}$ mode while keeping low porosity ($<$ 10\%). Further, to achieve elastic wave attenuation, different types of metamaterials have been studied; with local resonator~\cite{xiao2012flexural,zouari2018flexural}, embedded with scatterer ~\cite{yan2022propagation,huang2017multiple}, or acoustic black holes ~\cite{deng2021broad,tang2019periodic}.

In this study, we are interested in elastic band gaps formed by elastic waves passing through the solid due to dynamic excitations. The primary objective of the research reported in this paper is to generate the flexural band gaps in low-frequency regions (encompassing the first ten modes of the structure and typically in the 30-300 Hz range) without employing resonators or scatterers or secondary material. A detailed parametric investigation of various shapes shows that a cross shape with a high aspect ratio can generate low-frequency band gaps with low porosity of the unit cell metamaterial. The numerical predictions are validated with experiments on a finite plate with 3 $\times$ 3 unit cells.

The rest of the paper is organized as follows: the plate kinematics, the Bloch periodic boundary condition, and the finite element discretization are discussed in \Cref{sec:theoryform}. \Cref{sec:numexamp} presents the flexural band gaps obtained from the numerical study. The experimental setup and the comparison between the experiments and the numerical simulation is presented in \Cref{sec:experiprocedure}, followed by significant conclusions from this study in the last section.

\section{Theoretical formulation}
\label{sec:theoryform}
An isotropic homogeneous plate does not exhibit band gap characteristics, while a plate with a cutout introduces band gap~\cite{wang2011large}. Consider a plate with a cutout to understand the flexural band gap characteristics. The plate kinematics is based on the Reissner-Mindlin plate theory (RMPT), an extension of the Kirchhoff-Love plate theory. The RMPT is preferred over the latter, as it accommodates the use of $\mathcal{C}^0$ shape functions within the framework of the finite element method. This section presents a brief overview of the RMPT, followed by a description of spatial discretization and enforcement of Bloch periodicity.

\subsection{Basics of Reissner-Mindlin plate theory}
\label{theoform}
\begin{figure}[htpb]
\centering
\includegraphics[scale=0.5]{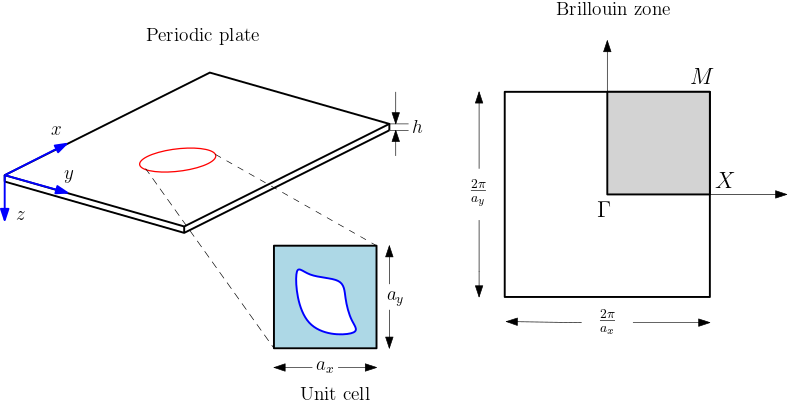}
\caption{Schematic of a periodic plate with uniform thickness, $h$ with its unit cell with an arbitrary cutout and the first irreducible Brillouin zone}
\label{fig:plategeo}
\end{figure}

Consider a mid-plane of an isotropic homogeneous elastic plate of uniform thickness, $h$, occupying the domain $\Omega \subset \mathbb{R}^2$, bounded by one-dimensional surface, $\partial \Omega$, see \Cref{fig:plategeo}. 
Let the coordinates of a point be denoted by $\xx = (x,y,z)$ and let the transverse displacement and the rotations normal to the midplane be denoted by $w(\xx)$ and $\boldsymbol{\beta}(\xx) = (\beta_x,\beta_y)$, respectively. The midplane displacements and the independent rotations within the RMPT are given by:
\begin{equation}
    \begin{aligned}
        u(\xx,t) &= z\beta_{x}(x,y,t) \\
        v(\xx,t) &= z\beta_{y}(x,y,t) \\
        w(\xx,t) &= w(x,y,t)
    \end{aligned}
    \label{u,v,w}
\end{equation}
The bending strain ($\bveps_{b}$) and the shear strain ($\bveps_s$) are given by:
\begin{equation}
    \bveps_{b} = 
    \begin{Bmatrix}   
        \dfrac{\partial \beta_{x}}{\partial x} \vspace{1mm}\\ 
         \dfrac{\partial \beta_{y}}{\partial y} \vspace{1mm}\\        
        \dfrac{\partial \beta_{x}}{\partial y} + \dfrac{\partial \beta_{y}}{\partial x} 
    \end{Bmatrix}; \quad
    \bveps_{s} = 
    \begin{Bmatrix}
        \gamma_{xz} \\
        \gamma_{yz} 
    \end{Bmatrix}  = 
    \begin{Bmatrix}
        \beta_{x} + \dfrac{\partial w}{\partial x} \vspace{1mm} \\
        \beta_{y} + \dfrac{\partial w}{\partial y} \\
    \end{Bmatrix}
    \label{transverseshearstrain}
\end{equation}
The bending and shear strains are related to the moment resultants, $\mathbf{M} = \{ M_{x}, M_{y}, M_{xy} \}$ and the shear forces $\mathbf{T} = \{ T_{x}, T_{y} \}$, respectively, are related through:
\begin{equation}
\mathbf{M}
    = \underbrace{\dfrac{E h^3}{12 (1 - \nu^2)}
\begin{bmatrix}
1 & \nu & 0 \\
\nu & 1 & 0 \\
0 & 0 & \dfrac{1 - \nu}{2}
\end{bmatrix}}_\mathbf{H_{b}} \bveps_{b}
    \quad \text{and} \quad
    \mathbf{T}
    = \underbrace{\kappa G h
\begin{bmatrix}
1 & 0 \\
 0 & 1 \\
\end{bmatrix}}_\mathbf{H_{s}} \bveps_{s}
    \label{moments and shearforce}
\end{equation}
where $\kappa$ is the shear correction factor, $G=E/(2(1+\nu))$ the bulk modulus, $E$ is the Young's modulus and Poisson's ratio is represented by $\nu$. In this paper, we adopt the Hellinger-Reissner function to develop the $n-$noded plate element and the total strain energy function, $\boldsymbol{\Pi}(\mathbf{u})$, is given by:
\begin{equation}
\Pi(\mathbf{u}) = {1 \over 2} \int_{\Omega} \left\{ \bveps_b^{\textup{T}} \mathbf{H}_{b} \bveps_b +  \bveps_s^{\textup{T}} \mathbf{H}_{s} \bveps_s \right\} \mathrm{d} \Omega - \int q~w~\rmd\Omega + \Pi_{\rm ext}
\label{eqn:potential}
\end{equation}
where $\mathbf{u} = \{w,\beta_x,\beta_y\}$ is the vector of the degrees-of-freedom (Dofs) associated with the displacement field in a finite element discretization; the influence of external forces and boundary is contained in $\Pi_{\rm ext}$ and $q$ is the transverse loading. The kinetic energy of the plate is given by:
\begin{equation}
T(\mathbf{u}) = {1 \over 2} \int_{\Omega} \left\{p (\dot{u}^2 + \dot{v}^2 + \dot{w}^2) + I(\dot{\beta}_x^2 + \dot{\beta}_y^2) \right\}~\mathrm{d} \Omega
\label{eqn:kinetic}
\end{equation}
where $\left(~\dot{}~\right)$, represents the time derivative of the variable, $p = \int_{-h/2}^{h/2} \rho(z)~dz, ~ I = \int_{-h/2}^{h/2} z^2 \rho(z)~dz$ and $\rho(z)$ is the mass density that varies through the thickness of the plate. In this work, the domain is partitioned into non-overlapping elements $\Omega^h$, and on using shape functions that span at least the linear space, the transverse displacement, $w$ and the independent rotations, $\beta_x,\beta_y$ are written as:
\begin{equation}
(w,\beta_x,\beta_y) = \sum_{i=1}^{n} \lambda_i \left(w_i , \beta_{x_i}, \beta_{y_i} \right) 
\label{eqn:apprx}
\end{equation} 
where $\lambda_i$ are the shape functions. For spatial discretiztion, the domain is discretized with 4-noded bilinear quadrilateral element with 3 dofs, viz., $(w,\beta_x,\beta_y)$ per node. To alleviate the shear locking phenomenon, a field consistent shear flexible element is used~\cite{somashekarprathap1987}. Upon substituting \Cref{eqn:apprx} in \Cref{eqn:potential,eqn:kinetic} and assuming harmonic solution $\mathbf{u} = \mathbf{u}_o e^{-i\omega t}$ for the unknown field is assumed and following standard Galerkin procedure, we get the following eigenvalue problem:
\begin{equation}
\left( \KK -\omega^2 \bfm \right) \mathbf{u}_o = \mathbf{0}
\label{eqn:freevibgovern}
\end{equation}
where $\mathbf{K} = \mathbf{K}_b + \mathbf{K}_s$ is the stiffness matrix that is an additive combination of the stiffness corresponding to the bending and shear, $\mathbf{M}$ is the mass matrix and $\omega$ is the frequency.


\subsection{Bloch Periodic boundary condition}
In this study, our focus is to generate flexural band gaps in the low frequency regime, wherein the domain kinematics is based on RMPT. Since this is a two-dimensional structure, and assuming the periodicity to be in the $x-$ and $y-$directions, the elastic waves propagate in the $xy$ plate, it should satisfy:
\begin{equation}
    \mathbf{u}(\mathbf{r}) = e^{i\mathbf{k}\cdot \mathbf{r}} \mathbf{u}_{\mathbf{k}}(\mathbf{r})
    \label{eqn:blochcond}
\end{equation}
where $\mathbf{k}$ is the first irreducible Brillouin zone, $\mathbf{u}_{\mathbf{k}}(\mathbf{r})$ is the vector function containing the degree-of-freedom and $\mathbf{r}$ is the position vector. 

\begin{figure}[htpb]
\centering 
\includegraphics[scale=0.7]{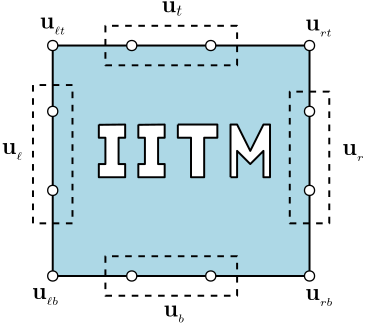}
\caption{Schematic representation of unit periodic cell and the notation for boundary nodes}
\label{fig:blockperiod}
\end{figure}

\Cref{fig:blockperiod} shows a representative unit cell. The nodes on the boundary are grouped into eight categories, viz., $\mathbf{u}_{_{\ell b}},~\mathbf{u}_{_b},~\mathbf{u}_{_{r b}}~\mathbf{u}_{_r},~\mathbf{u}_{_{r t}}.~\mathbf{u}_{_{t}},~\mathbf{u}_{_{\ell t}}$ and $\mathbf{u}_{_{\ell}}$, where the subscripts $\ell,~r,~b,~t$ represents the displacements corresponding to the left, right, bottom and top nodes of the unit cell. Note that the corner nodes are represented with a double subscript. All the interior nodes are grouped in $\mathbf{u}_{_{in}}$. Using \Cref{eqn:blochcond}, the displacement relations between the about categories is given by:
\begin{equation}
\begin{aligned}
    \mathbf{u}_{_r} &= e^{i k_x a}\mathbf{u}_{_\ell}, \qquad \mathbf{u}_{_t} = e^{i k_y a}\mathbf{u}_{_b} \\
    \mathbf{u}_{_{r b}} &= e^{i k_x a} \mathbf{u}_{_{\ell b}}, \qquad \mathbf{u}_{_{r t}} = e^{i (k_x + k_y)a} \mathbf{u}_{_{\ell b}} \\
    \mathbf{u}_{_{\ell t}} &= e^{i k_y a} \mathbf{u}_{_{\ell b}}.
\end{aligned}
\end{equation}
According to the Bloch periodicity, the displacements in the domain can be related to the above eight categories through the following transformation matrix:
\begin{equation}
    \mathbf{u}_o = \mathbf{T} \hat{\mathbf{u}}
    \label{eqn:displacement_transformation}
\end{equation}
where $\hat{\mathbf{u}} = \left\{\mathbf{u}_{_\ell},~\mathbf{u}_{_b},~\mathbf{u}_{_\ell b},~\mathbf{u}_{_{in}} \right\}^{\rm T}$ and the transformation matrix $\mathbf{T}$ is given by:
\begin{equation}
    \mathbf{T} = \begin{bmatrix} \mathbf{I} & \mathbf{0} & \mathbf{0} & \mathbf{0} \\
    \mathbf{I}e^{k_x a} & \mathbf{0} & \mathbf{0} & \mathbf{0} \\
    \mathbf{0} & \mathbf{I} & \mathbf{0} & \mathbf{0} \\
    \mathbf{0} & \mathbf{I}e^{k_ya} & \mathbf{0} & \mathbf{0} \\
    \mathbf{0} & \mathbf{I} & \mathbf{I} & \mathbf{0} \\
    \mathbf{0} & \mathbf{I} & \mathbf{I}e^{k_xa} & \mathbf{0} \\
    \mathbf{0} & \mathbf{I} & \mathbf{I}e^{k_ya} & \mathbf{0} \\
    \mathbf{0} & \mathbf{I} & \mathbf{I}e^{(k_x+k_y)a} & \mathbf{0} \\
    \mathbf{0} & \mathbf{I} & \mathbf{0} & \mathbf{I} 
    \end{bmatrix}
\end{equation}
Upon substituting \Cref{eqn:displacement_transformation} into \Cref{eqn:freevibgovern}, results in:
\begin{equation}
 \mathbf{A}(k_x,k_y,\omega) \hat{\mathbf{u}} = \mathbf{0}
 \label{eqn:blochwave}
\end{equation}
where $\mathbf{A}(k_x,k_y,\omega) = \Big( \mathbf{T}^{\dagger} (\mathbf{K} - \omega^2 \mathbf{M}) \mathbf{T} \Big)$ and `$\dagger$' denotes the Hermitian transpose. By solving \Cref{eqn:blochwave}, the eigen frequencies corresponding to a specific Bloch wave vector $(k_x,k_y)$ can be computed. This is repeated for all Bloch wave vectors along the path $M - \Gamma - X - M$ (c.f. \Cref{fig:plategeo}), a band diagram can be constructed to determine the dispersion spectra of the considered unit cell.

\section{Numerical Results and Discussion}
\label{sec:numexamp}
\subsection{Validation}
\begin{figure}[htpb]
    \centering
    \includegraphics[width= 12cm]{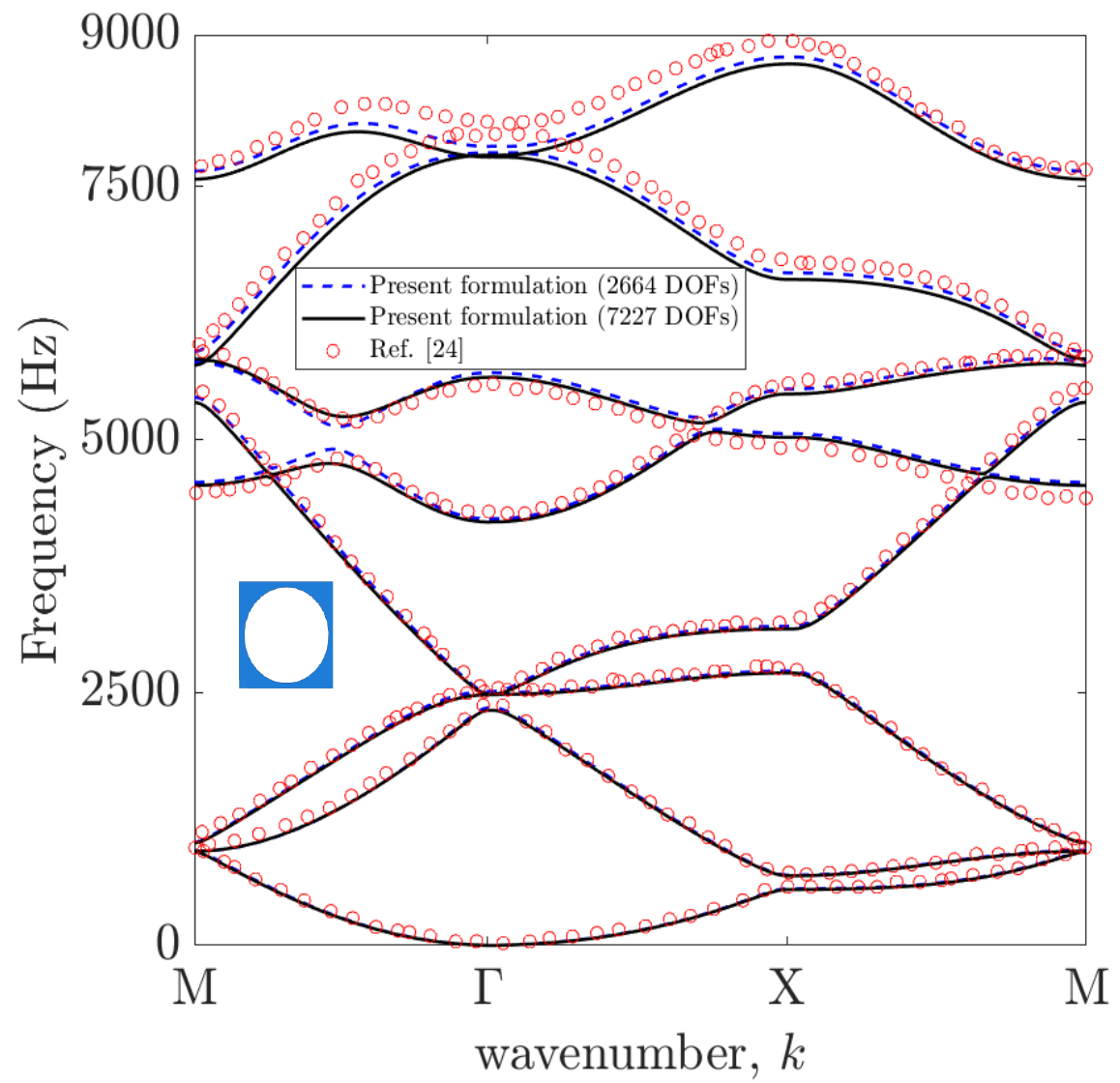}
    \caption{Comparison of band gaps obtained from the present work with that of Huang  et al. \cite{huang2018multiple}.}
    \label{fig:va}
\end{figure}

To validate the accuracy of the present method, the results were compared with those from Huang et al.~\cite{huang2018multiple}. For this purpose, we consider a plate of size $a=$ 10 cm, $h=$ 1 cm with a cutout radius, $r=$ 4.5 cm. The material is assumed to be homogeneous and isotropic with the following properties: density, $\rho=$ 1144.4 kg/m$^3$, Youngs' modulus, $E =$ 3.228 GPa and Poisson’s ratio, $\nu =$ 0.37. The computational domain is discretized with 4-noded bilinear quadrilateral elements, care has been taken to have the same number of nodes on the boundary that facilitates the application of Bloch periodic boundary conditions. Based on a systematic study, a mesh consisting of 7227 degrees of freedom was found to be adequate to compute the dispersion curves. \Cref{fig:va} shows the dispersion curves for a square plate with a circular cutout, and the results from the present work are compared against the work of Huang et al.,~\cite{huang2018multiple}. Note that the authors have not mentioned the mesh size for their results. The simulation was also run on a relatively coarser mesh with 2664 degrees of freedom, and it is seen that the results from the present work are in close agreement with that of Huang et al.,~\cite{huang2018multiple}.


\subsection{Effect of shape, size and location of cutouts}
Many studies have focused on the porous metamaterial with a circular hole~\cite{wang2011large,huang2018multiple}. However, to the best of the authors' knowledge, the influence of the cutout's shape, size, and location on the width and the number of band gaps have not been studied systematically. As mentioned earlier, this study focuses on band gaps in the lower frequency zones (within the first 3 to 4 modes). Hence, this study investigated a series of shapes, starting with an ellipse and progressively moving to slender shapes with higher aspect ratios to understand their effect on the generation of band gaps in the low-frequency regime.

\begin{figure}[htpb]
    \centering
    \begin{subfigure}{0.3\textwidth}
        \centering
        \includegraphics[width=\textwidth]{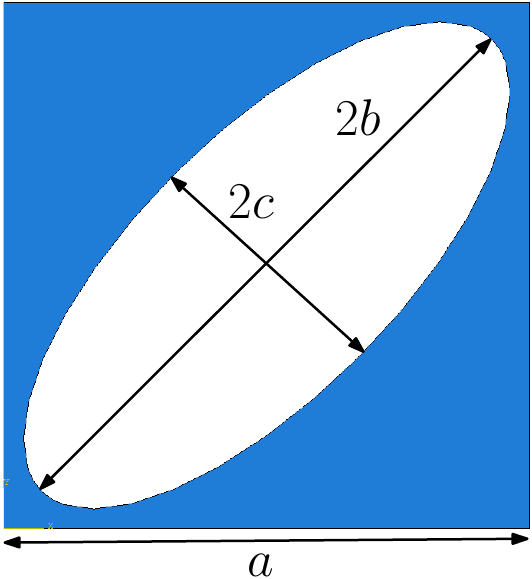}
        \caption{}
        \label{fig:1ellipse}
    \end{subfigure}
    \hfill
    \begin{subfigure}{0.3\textwidth}
        \centering
        \includegraphics[width=\textwidth]{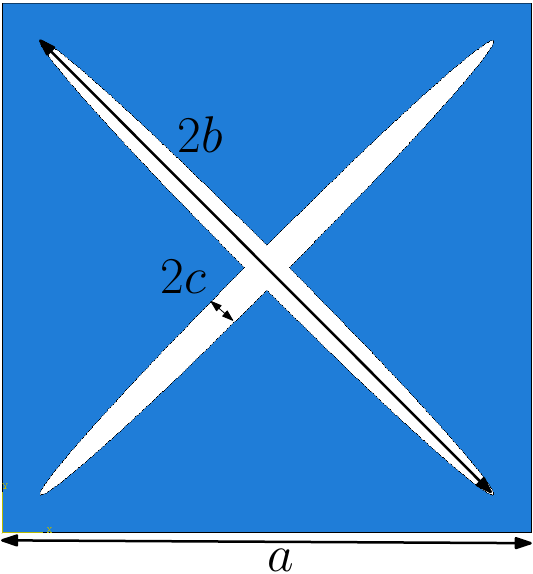}\\
        \caption{}
        \label{fig:2DigEllp}
    \end{subfigure}   
 \hfill
    \begin{subfigure}{0.3\textwidth}
        \centering
        \includegraphics[width=\textwidth]{Images/Unit_cell_1DigElp.pdf}\\
        \caption{}
        \label{fig:2DigRect}
    \end{subfigure} 
 \hfill
    \begin{subfigure}{0.3\textwidth}
        \centering
        \includegraphics[width=\textwidth]{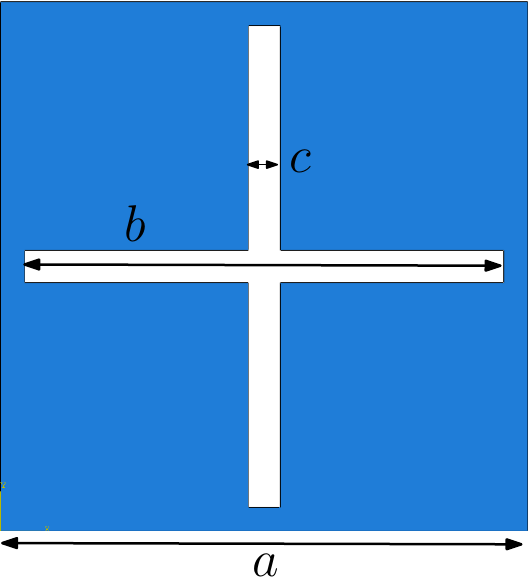}\\
        \caption{}
        \label{fig:PlusSpahe}
    \end{subfigure} 
    \hfill
    \begin{subfigure}{0.3\textwidth}
        \centering
        \includegraphics[width=\textwidth]{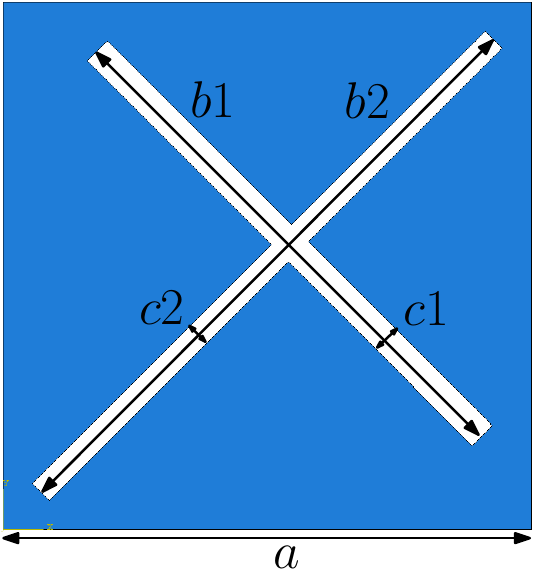}\\
        \caption{}
        \label{fig:1DigMoved}
    \end{subfigure} 
    \hfill
    \begin{subfigure}{0.3\textwidth}
        \centering
        \includegraphics[width=\textwidth]{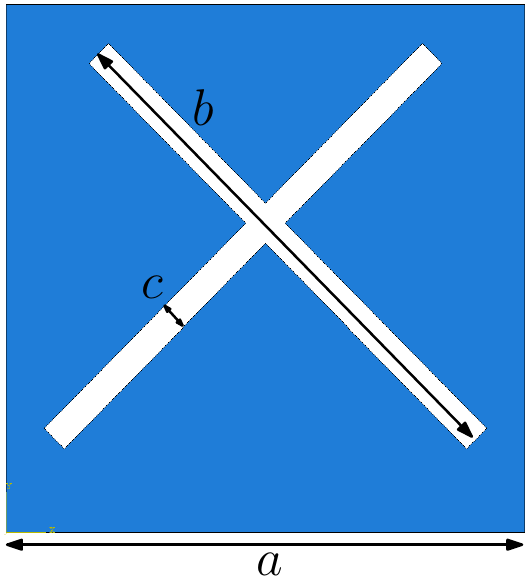}\\
        \caption{}
        \label{fig:2DigMoved}
    \end{subfigure} 
    \caption{Schematic representation of a unit cell with different cutout shapes: (a) ellipse, (b) intersecting ellipses, (c) intersecting rectangular slots, (d) cross, (e) unequal intersecting slots with centre offset and (f) Intersecting slots with centre offset.}
    \label{fig:Unitcells}
\end{figure}

For the numerical study, the material used is Aluminum with Young's modulus, $E =$ 70 GPa, density, $\rho =$ 2700 kg/m$^3$, and Poisson's ratio, $\nu =$ 0.33. We consider a square unit cell with side length, $a=$ 330 mm, and thickness, $h=$ 2 mm. Six different configurations are considered to study the influence of cutout shape, size and location. \Cref{fig:Unitcells} shows the unit cells with different cutout shapes and the dimensions of the cutouts are given in \Cref{tab:cutout_dimension}. Note that the dimensions for the cutouts, c.f. \Cref{fig:2DigEllp,fig:2DigMoved} are arrived such that the area removed is the same, i.e., about 11$\%$ area is removed.
\begin{table}[htpb]
\centering
\caption{Unit cell with different configurations}
    \begin{tabular}{ll}
    \hline
    Cavity shape & dimensions (in mm) \\
    \hline
    Ellipse & $b=$ 200, $c=$ 100  \\
    Intersection ellipses &  $b=$ 200, $c=$ 10  \\
    Intersection rectangular slots & $b=$ 400, $c=$ 15 \\
    Cross & $b=$ 300, $c=$ 20 \\
    Unequal intersecting slots with centre offset & $b1=$ 340, $c1=$ 17.72 \\
    & $b2=$ 400, $c2=$ 15 \\
    Intersecting rectangular slots with centre offset & $b=$ 340, $c=$ 17.8\\
         \hline
    \end{tabular}
\label{tab:cutout_dimension}
\end{table}

\begin{figure}[htpb]
    \centering
    \begin{subfigure}{0.32\textwidth}
        \centering
        \includegraphics[width=\textwidth]{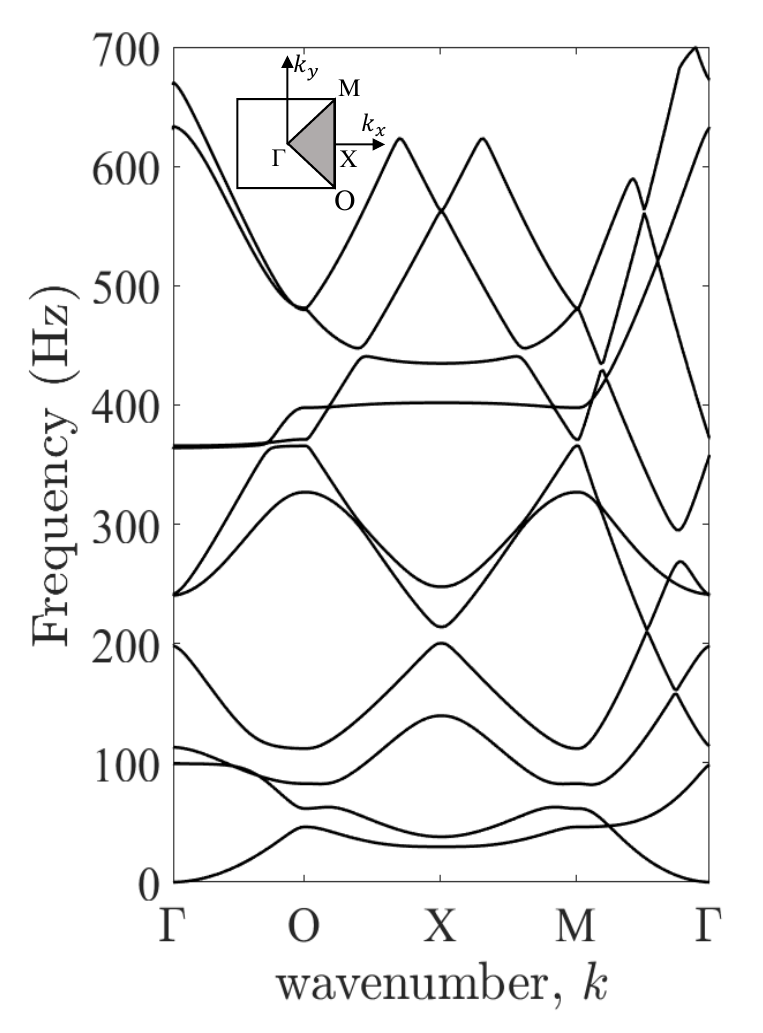}
        \caption{}
        \label{fig:1Elps}
    \end{subfigure}
    \hfill
    \begin{subfigure}{0.3\textwidth}
        \centering
        \includegraphics[width=\textwidth]{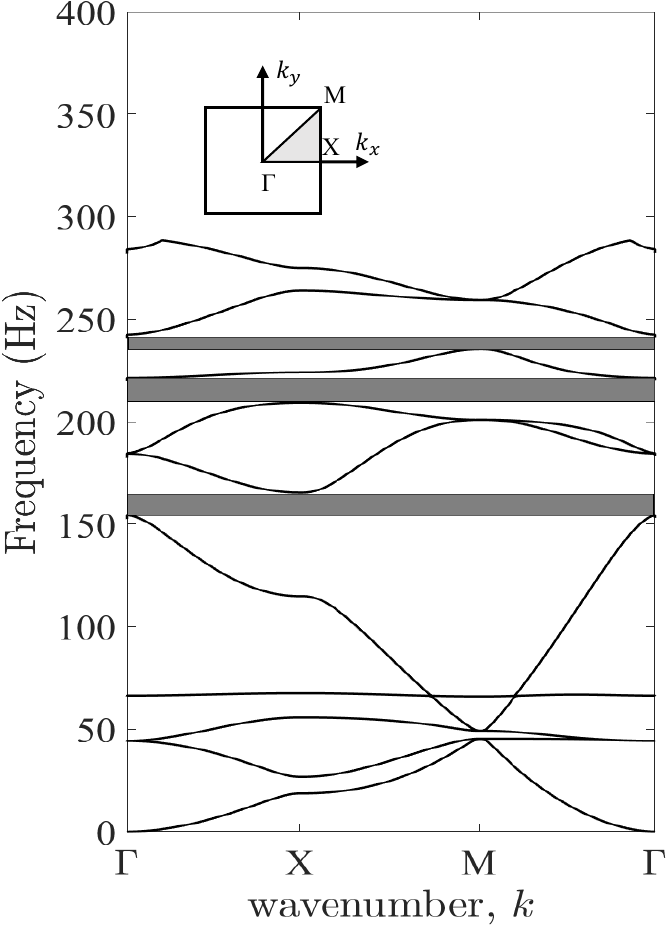}\\
        \caption{}
        \label{fig:2DigElp}
    \end{subfigure}   
 \hfill
    \begin{subfigure}{0.3\textwidth}
        \centering
        \includegraphics[width=\textwidth]{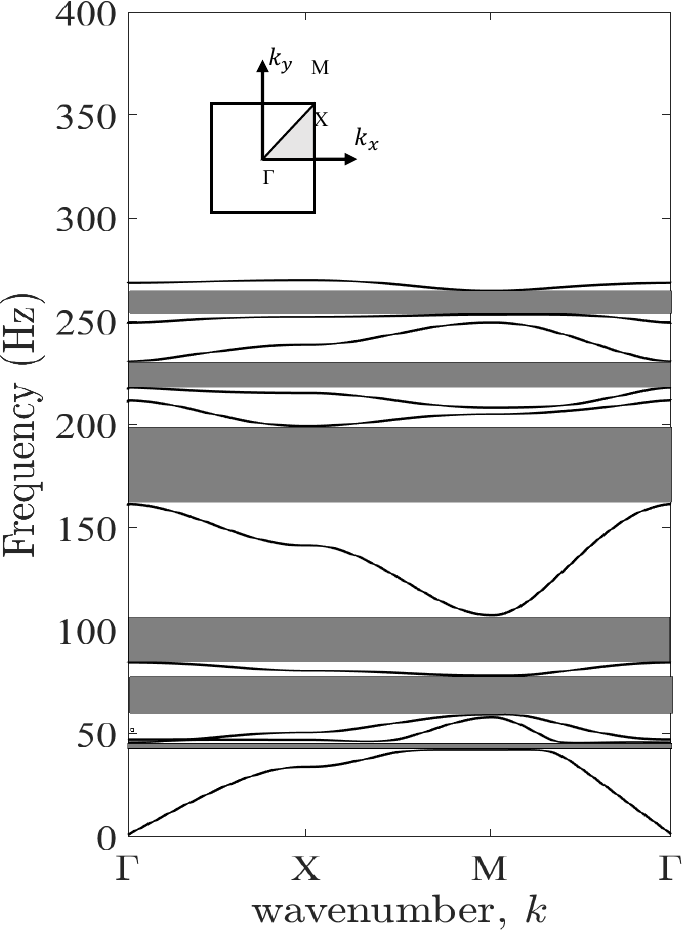}
        \caption{}
        \label{fig:2DigRectB}
    \end{subfigure} 
 \hfill
    \begin{subfigure}{0.3\textwidth}
        \centering
        \includegraphics[width=\textwidth]{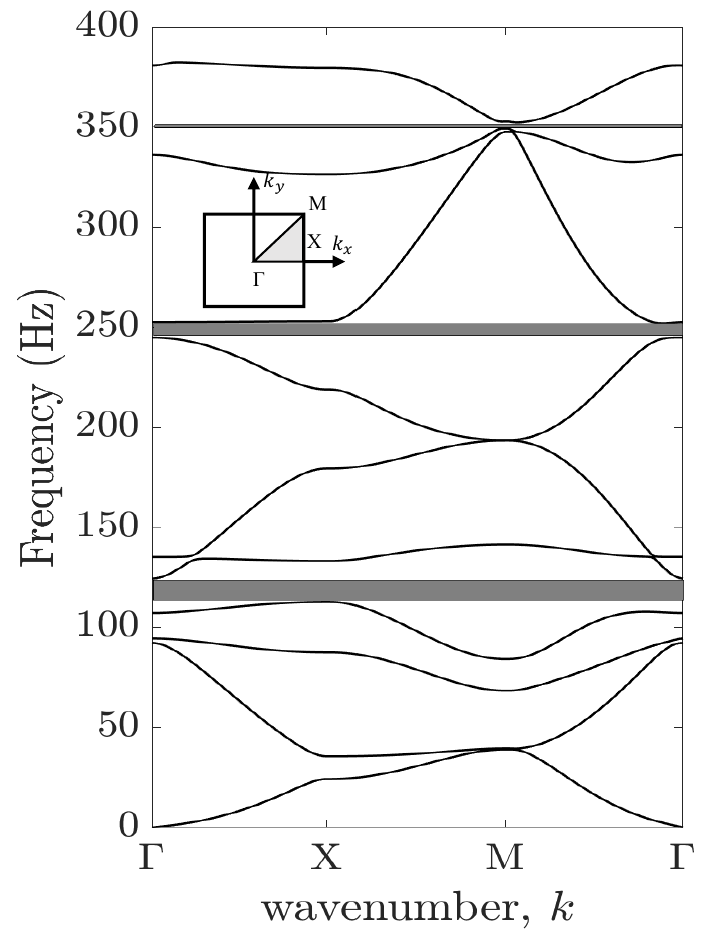}
        \caption{}
        \label{fig:PlusShp}
    \end{subfigure}    
\hfill
    \begin{subfigure}{0.3\textwidth}
        \centering
        \includegraphics[width=\textwidth]{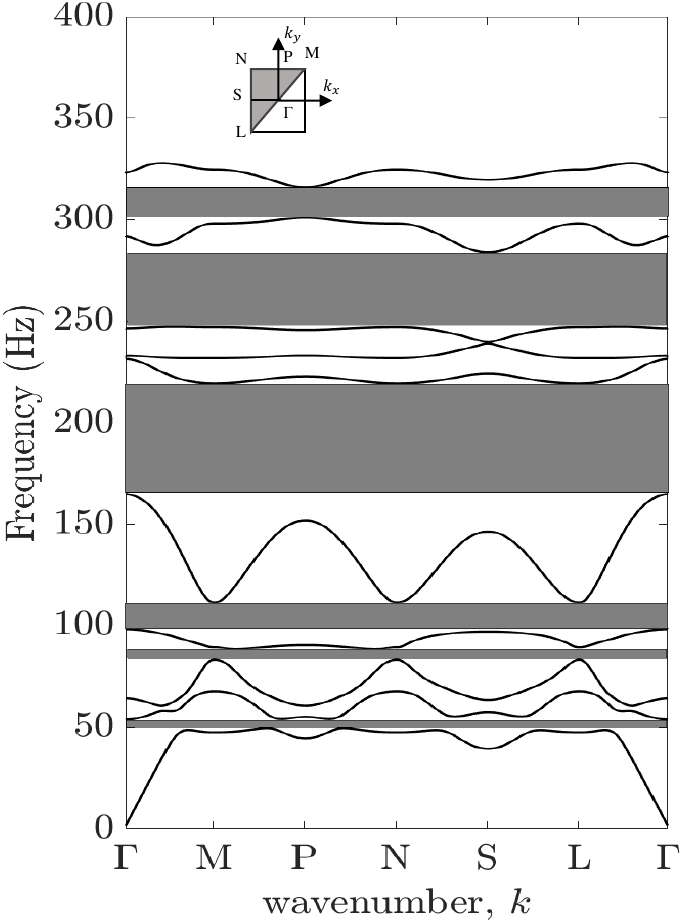}
        \caption{}
        \label{fig:1DigM}
    \end{subfigure}
\hfill
    \begin{subfigure}{0.3\textwidth}
        \centering
        \includegraphics[width=\textwidth]{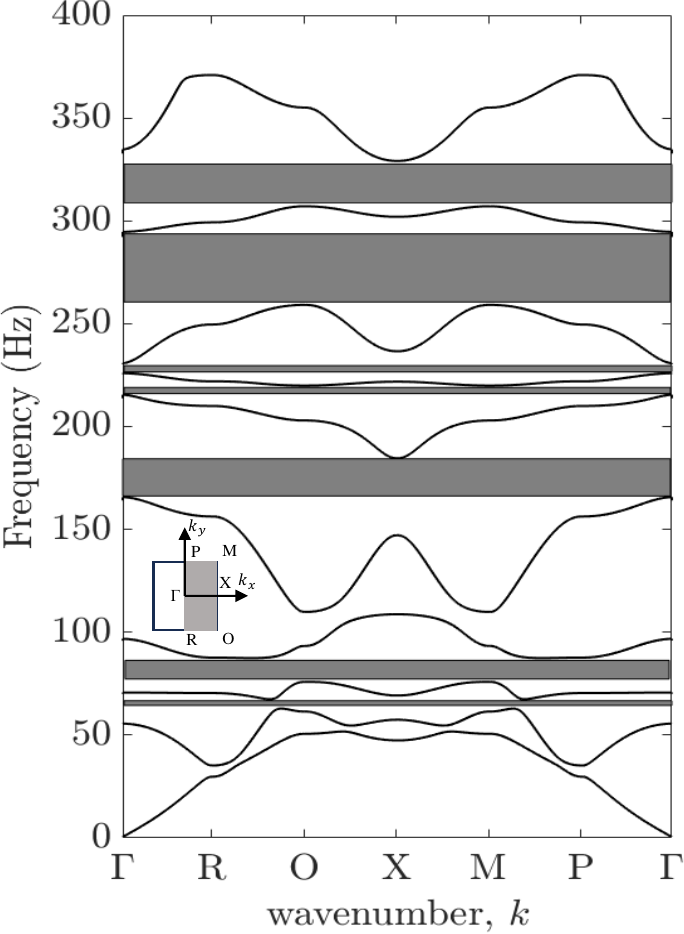}
        \caption{}
        \label{fig:2DigM}
    \end{subfigure}
\caption{Dispersion curves for the unit cell for a plate with various cutouts: (a) ellipse, (b) intersecting ellipses, (c) intersecting rectangular slots, (d) cross, (e) unequal intersecting slots with centre offset and (f) Intersecting slots with centre offset.}
    \label{fig:Bandss}
\end{figure}

\Cref{fig:Bandss} shows the dispersion curves for the considered cutout configurations. It seen from \Cref{fig:1Elps} that for an elliptical cutout, no band gaps are present. Following this, the single ellipse is replaced by two high aspect ratio slender elliptical slots with their semi-major axes at 90 degrees (c.f. \Cref{fig:2DigEllp}). From \Cref{fig:2DigElp}, it can be seen that three band gaps are present between the $5^{\rm th}-6^{\rm th}$, $7^{\rm th}-8^{\rm th}$ and $8^{\rm th}-9^{\rm th}$ modes. In the case of intersecting rectangular slots, in addition to the band gaps that is present for the intersecting ellipses, band gaps also appear between the following modes: $1^{\rm st}-2^{\rm nd}$, $3^{\rm rd}-4^{\rm th}$, $4^{\rm th}-5^{\rm th}$. The maximum bandwidth of 37.86 Hz is observed between $5^{\rm th}-6^{\rm th}$, followed by the bandwidth of 19.02 Hz and 23.20 Hz between the 3$^{\rm rd}-4^{\rm th}$ and $4^{\rm th}-5^{\rm th}$, respectively.


 
It appears that removing material towards the corners and at the plate's center affects the band gaps' presence and size. Two more geometries are generated as in \Cref{fig:1DigMoved,fig:2DigMoved} to understand this. In the first case, one of the rectangular slots is moved away from the center. So the area removed around the center has now been reduced. From \Cref{fig:1DigM} multiple wide band gaps can be seen, but the band gaps below 100 Hz became narrower and band above 150 Hz becomes wider. In the following study, both the rectangular holes are moved away from the center while keeping the porosity the same. \Cref{fig:2DigM} shows the dispersion curves for both the slots being offset from the center; it can be seen that the bands became narrower, and there are two narrow bands below 100 Hz. This would suggest that more material needs to be removed around the center area to generate wider band gaps in the lower frequencies (50--150 Hz).

The results imply that band gap generation is \emph{highly} sensitive to the geometric shape and size of the cutouts. With low porosity and high aspect ratio, rectangular slots (along the diagonal and passing through the center of the plate) can generate multiple band gaps in the lower modes (first four). The porosity obtained with these diagonal slots is about 11\%, rendering it a practical solution for vibration isolation applications targeting low frequencies.

\section{Experimental Verification}
\label{sec:experiprocedure}

\subsection{Experimental setup}
To verify the numerical results, an experiment is carried out to examine the attenuation capabilities of the proposed low-porosity perforated plate. An aluminum square plate with a side of 1070 mm is considered. The side length of 1070 mm corresponds to 9 unit cells, with each unit cell having an `x' shape cutout(c.f. \Cref{fig:2DigRect}); the cutouts are manufactured using laser cutting. To demonstrate the effectiveness of this cutout on vibration isolation, the experiments and the FE simulations were carried out on a solid aluminum plate as well.



\begin{figure}[htpb]
    \centering
    \begin{subfigure}{\textwidth}
        \centering
        \includegraphics[width = 10cm]{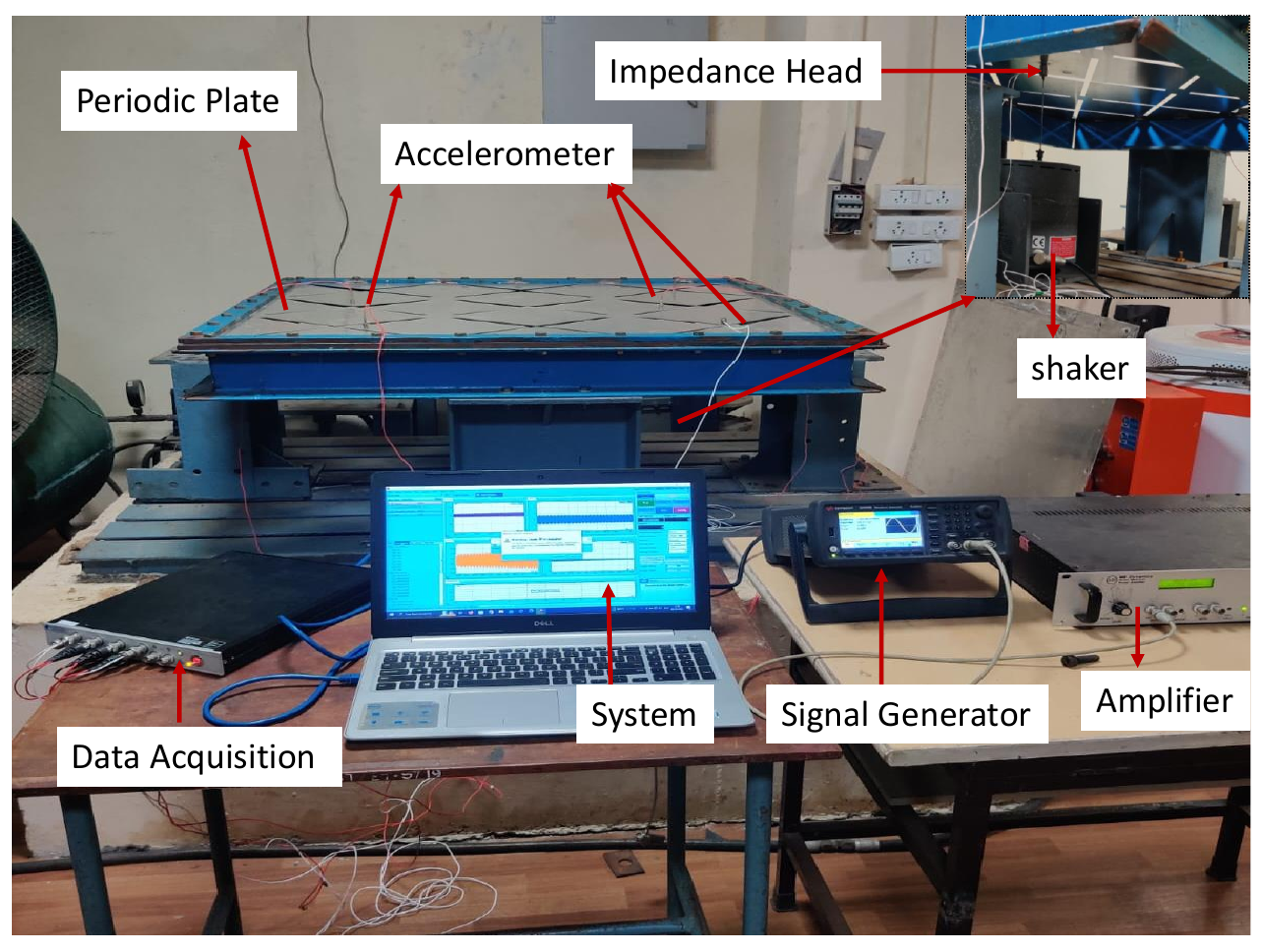}
        \caption{}
        \label{fig:Exp setp}
    \end{subfigure}
    \hfill
    \begin{subfigure}{\textwidth}
        \centering
        \includegraphics[width = 10cm]{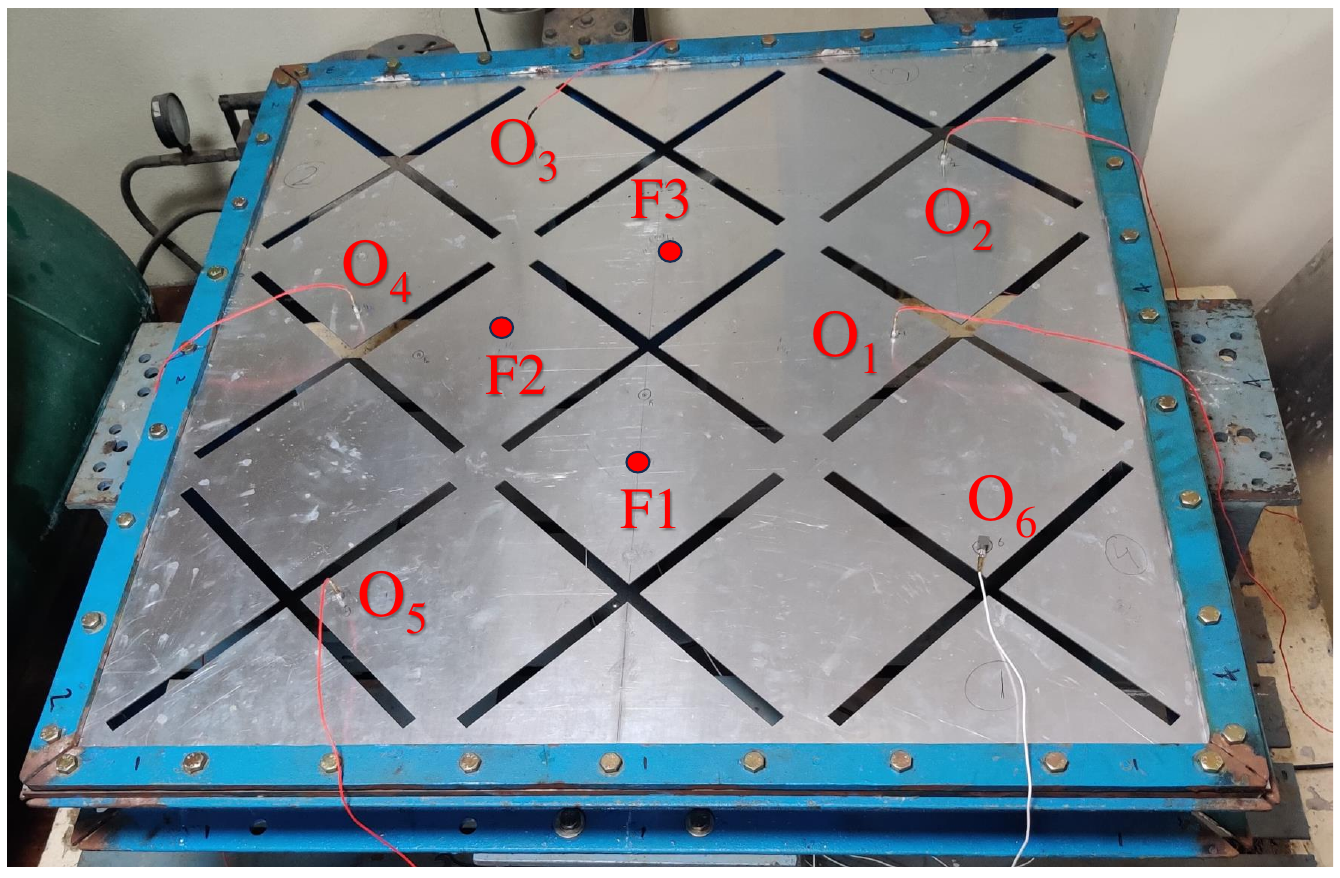}\\
        \caption{}
        \label{fig:location}
    \end{subfigure}   
     \caption{(a)Experiment Setup, (b) Plate with locations of input force and response measurement.}
    \label{fig:Exp}
\end{figure}

\Cref{fig:Exp setp} shows the experimental setup. The plate is clamped on all sides, and the shaker (MB Dynamics) is attached at the bottom of the plate. The signal generator (Keysight waveform generator) provides excitation to the shaker through an amplifier. The shaker tip has an impedance head  (Dytran model 5860B) to measure the input force and acceleration, and accelerometers at multiple locations (Dytran model 3145AG) are used to measure the response. An 8-channel data acquisition system (Crystal Instruments) is used to acquire and process sensor data.

For the experiment, a sine sweep signal from 30 to 300 Hz is used for exciting the plate structure. The shaker is moved to three locations, F1, F2, and F3, as shown in \Cref{fig:location}; this is done to understand the effect of excitation location on the plate's response. The shaker is placed at one of the three locations, and the response is measured at 6 locations ($O_i,~i=1,\cdots,6$), as shown in \Cref{fig:location}. Four averages are used at each accelerometer measurement location to reduce signal noise. A composite root mean square (RMS) acceleration is obtained from the accelerations measured at the various locations as $a_{RMS} = \sqrt{\frac{\Sigma_{i=1}^N a_i^2}{N}}$; in our case $N=6$. This is normalized with the RMS value of the acceleration at the excitation location. We do this so that a single metric can be used to compare the effect of excitation location and the effect of cut-outs at different frequencies.

\subsection{Results}
Next, the plates' normalized response from FEM is compared with the experiment, using the same six measurement locations $O_i,~i=1,\cdots,6$ for different excitation locations, $F_i,~i=1,2,3$. A comparison between the periodic plate and the finite plate is also presented. In the forthcoming figures, the band gap obtained from the FE simulation on the periodic plate is highlighted with a `grey' region.

\begin{figure}[htpb]
 \centering
  \begin{subfigure}{\textwidth}
        \centering
        \includegraphics[height = 7cm]{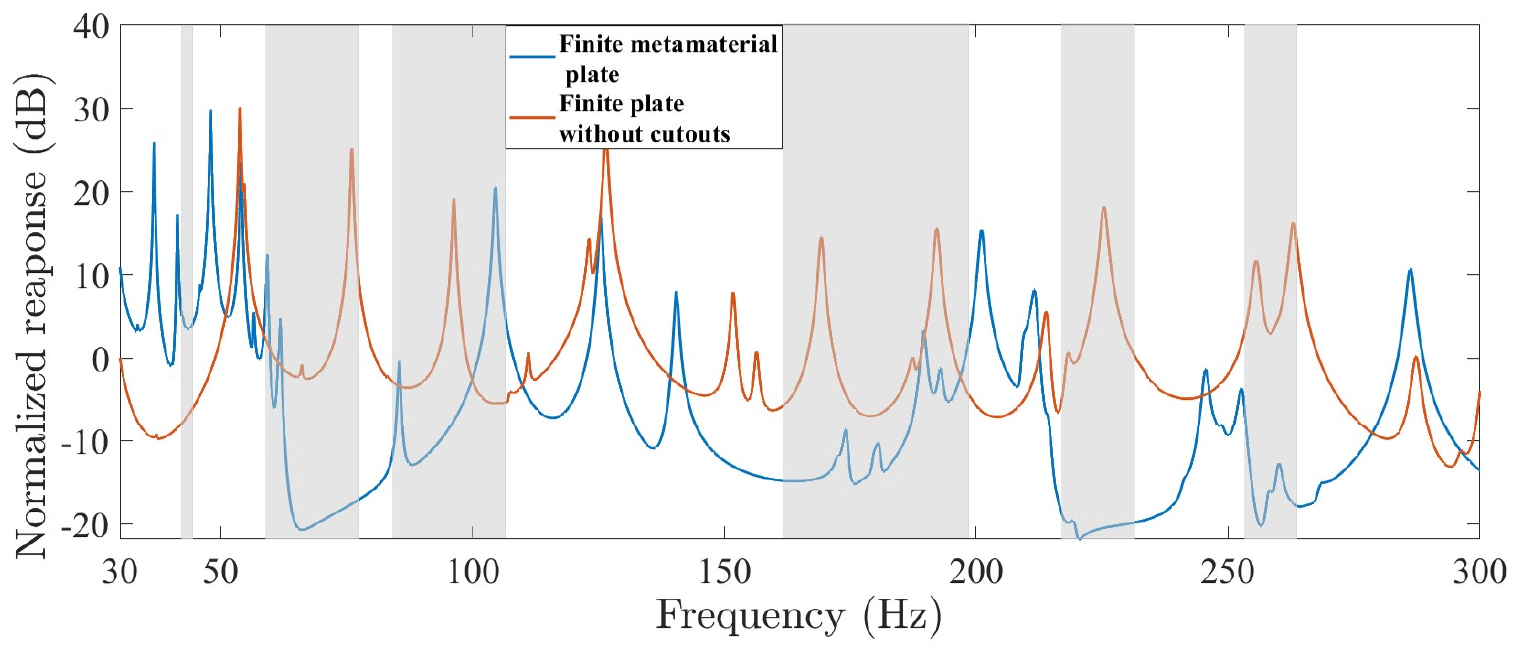}
        \caption{}
        \label{fig:FEM_F1}
  \end{subfigure}
\hfill
    \begin{subfigure}{\textwidth}
        \centering
        \includegraphics[height = 7cm]{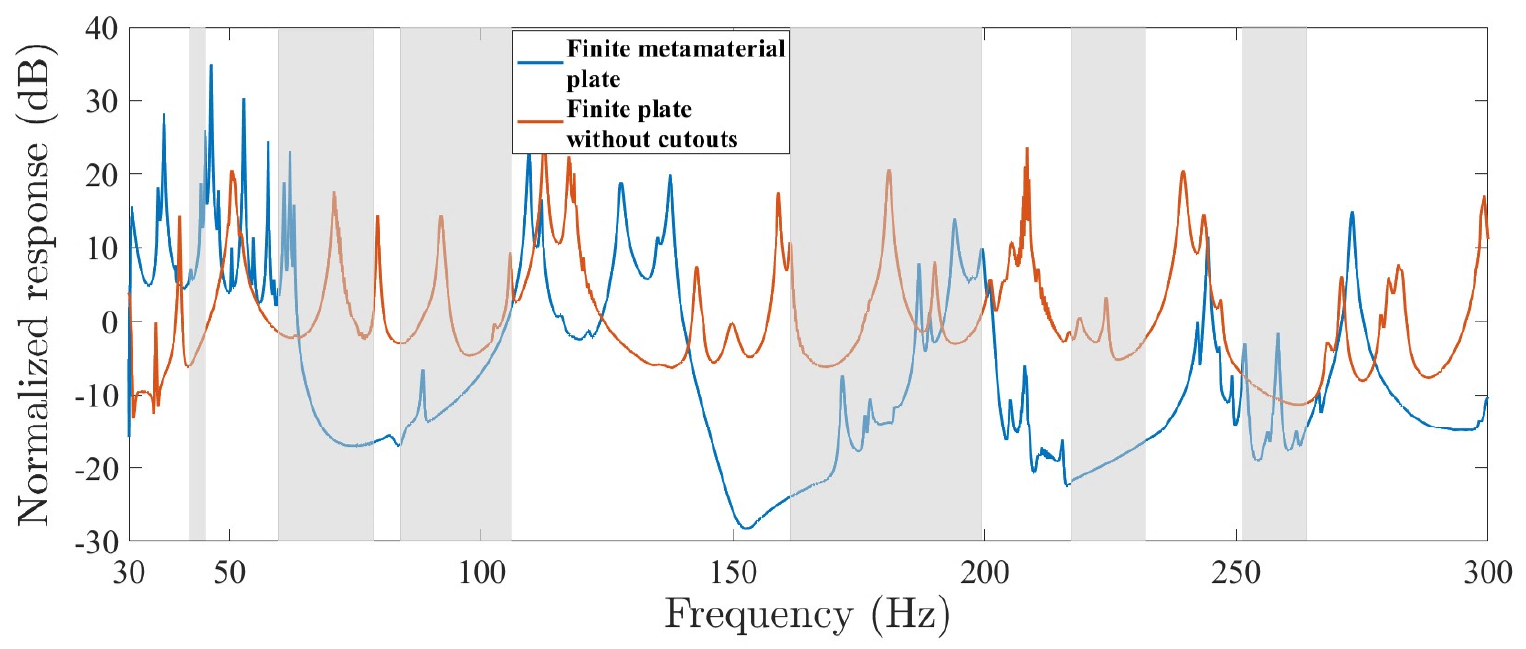}
        \caption{}
        \label{fig:F1}
    \end{subfigure}
    \caption{Normalized Response for input at F1 (a)FEM (b) Experiment}
    \label{fig:F1resp}
\end{figure}

For the excitation location F1, \Cref{fig:FEM_F1,fig:F1} shows a comparison between the periodic plate and the solid plate normalized response obtained from FEM and experiment. The shaded regions represent the band gaps from the numerical study of the infinite periodic structure. It can be seen that the periodic plate response is lower than the solid plate for all the bands except in the first band gap region. The numerical simulations show that Bands 2 and 3 are smaller than the bandwidth in the unit cell model. The experimental results also show the same trend, but the 2$^{\rm nd}$ and 4$^{\rm th}$ frequency bands are smaller than the bandwidth the unit cell model predicted. The 5$^{\rm th}$ and 6$^{\rm th}$ band of the FEM and experiment response shows good agreement with the numerical band. Overall, one can see significant vibration reduction in the stop band regions both in FE simulations and experiments. 

\begin{figure}[htpb]
 \centering
  \begin{subfigure}{\textwidth}
        \centering
        \includegraphics[height = 6cm]{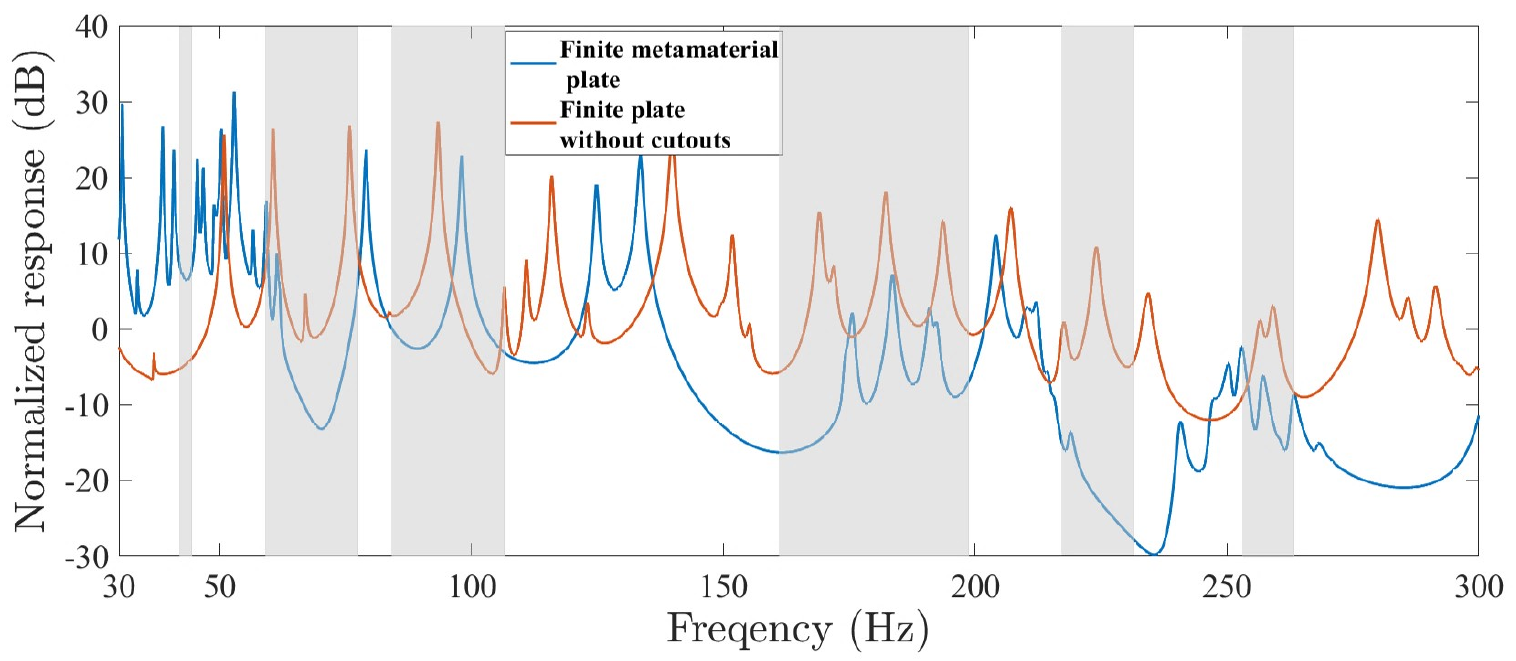}
        \caption{}
        \label{fig:FEM_F2}
  \end{subfigure}
\hfill
    \begin{subfigure}{\textwidth}
        \centering
        \includegraphics[height = 6cm]{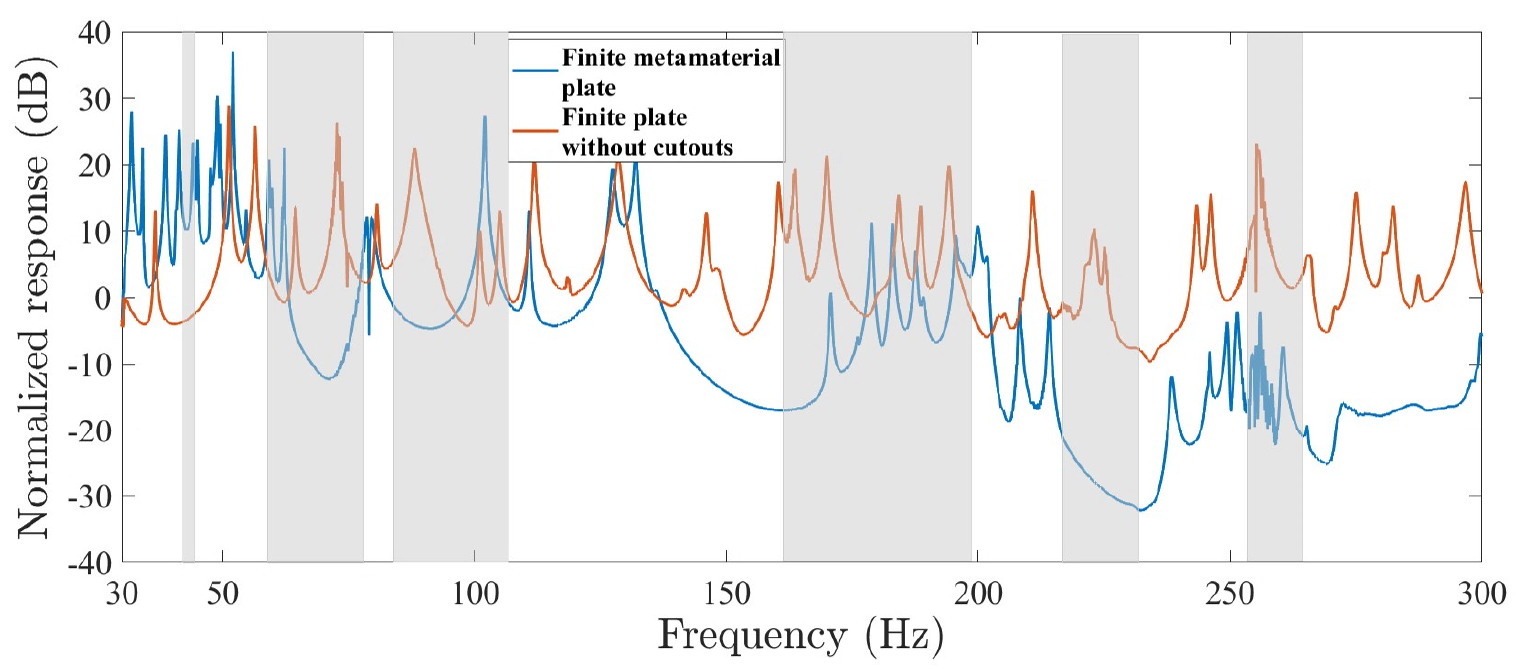}
        \caption{}
        \label{fig:F2}
    \end{subfigure}
    \caption{Normalized Response for input at F2 (a)FEM (b) Experiment}
    \label{fig:F2resp}
\end{figure}

For the excitation location F2, \Cref{fig:FEM_F2,fig:F2} shows a comparison between the periodic plate and the solid plate normalized response obtained from FEM and experiment. The shaded regions again represent band gaps from the numerical study of the infinite structure. It can be seen that the FEM and experiment are consistent with each other. In the first band gap region, both FEM and experiment show higher response for the periodic plate when compared to the solid plate. In the 2$^{\rm nd}$ band, the response of the periodic plate is lower than the solid plate, but the frequency band is smaller than the numerical band gap. For the third numerical band, the periodic plate shows not much improvement in both FEM and experiment. In the experiment, the periodic plate response is lower but for a small frequency range in the 4$^{\rm th}$ numerical band, while in FEM, the periodic plate response is lower in the complete band. Both FEM and experiment show good agreement with the numerical band (5$^{\rm th}$ and 6$^{\rm th}$).

\begin{figure}[htpb]
 \centering
  \begin{subfigure}{\textwidth}
        \centering
        \includegraphics[height = 6cm]{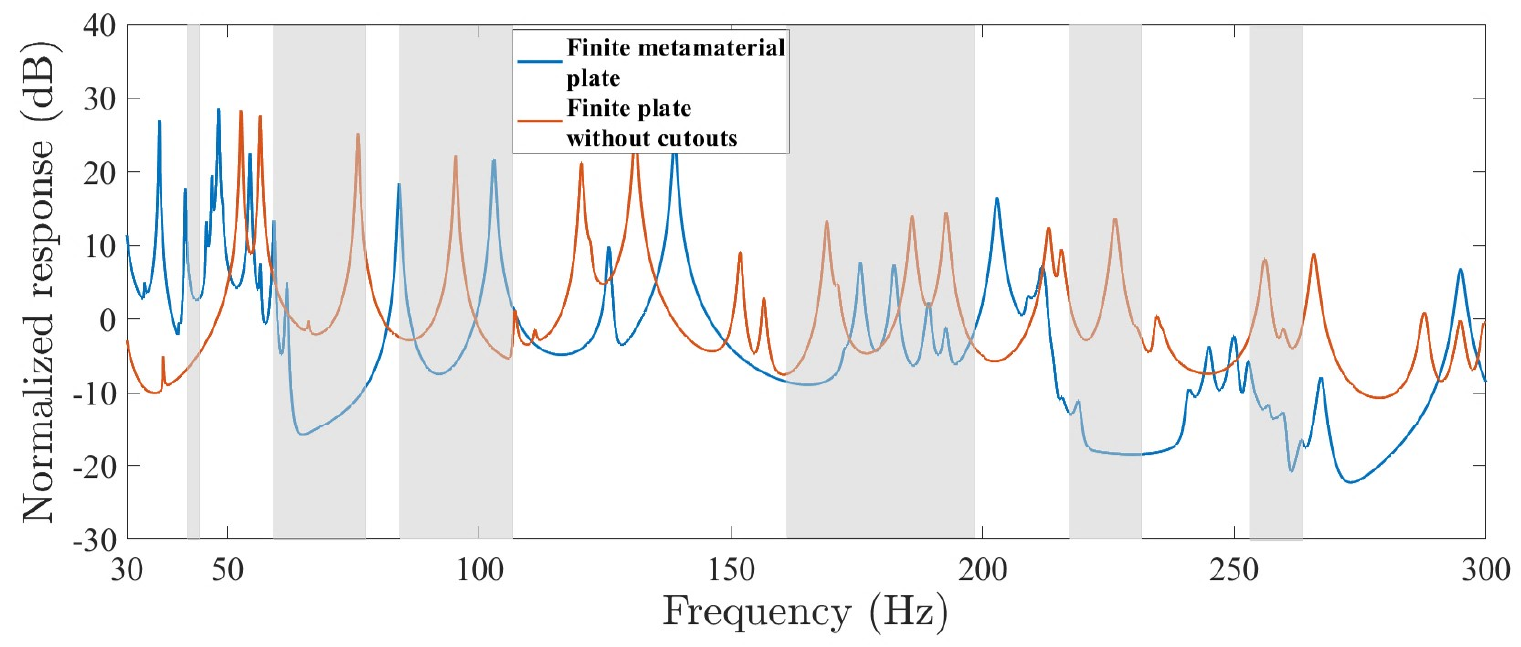}
        \caption{}
        \label{fig:FEM_F3}
  \end{subfigure}
\hfill
    \begin{subfigure}{\textwidth}
        \centering
        \includegraphics[height = 6cm]{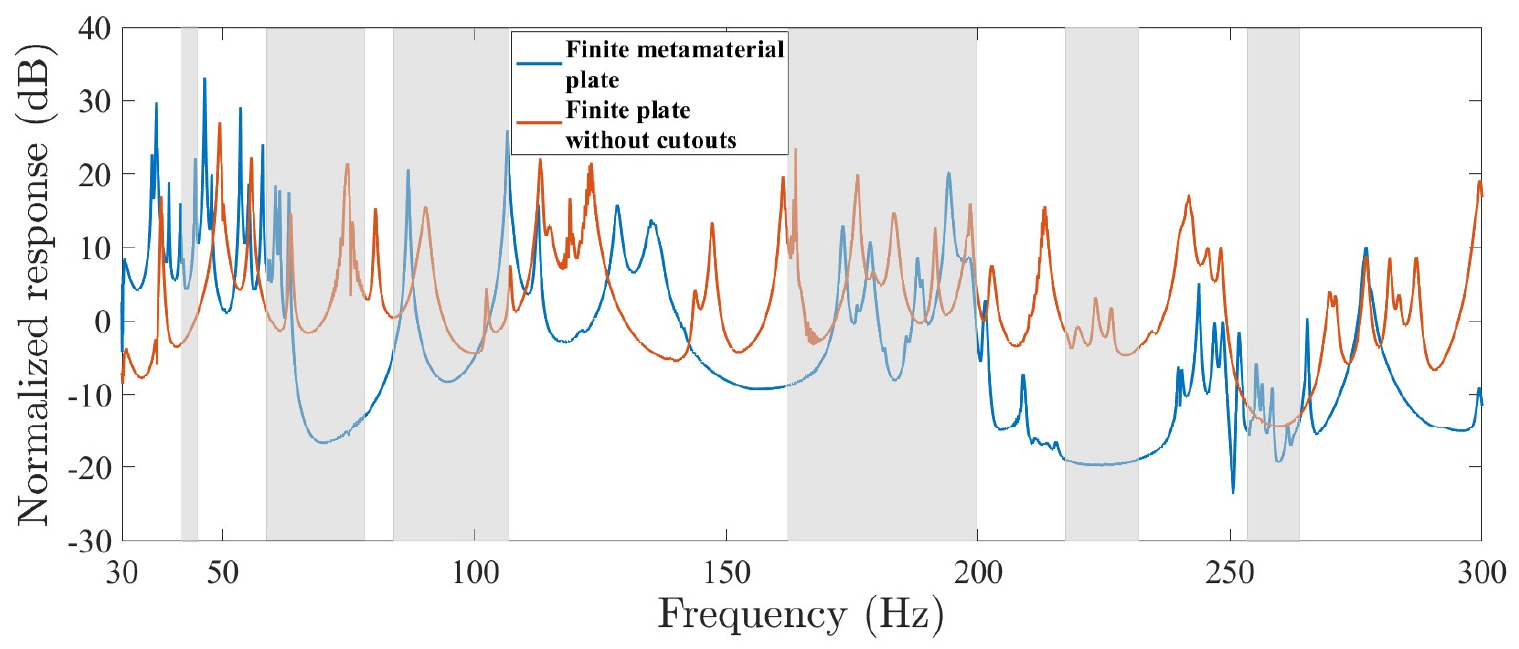}
        \caption{}
        \label{fig:F3}
    \end{subfigure}
    \caption{Normalized Response for input at F3 (a)FEM (b) Experiment}
    \label{fig:F3resp}
\end{figure}

For the excitation location F3, \Cref{fig:FEM_F3,fig:F3} shows a comparison between the periodic plate and the solid plate normalized response obtained from FEM and experiment.  In 1st band, both FEM and experiment show periodic plate response higher than solid plate. The frequency range for which periodic plate response is smaller than solid plate is smaller than the 2$^{\rm nd}$ numerical band in both FEM and experiment. In the 3$^{\rm rd}$ and 4$^{\rm th}$ bands, the periodic plate does not show much of an improvement over the solid plate in both FEM and experiment. FEM and experiment response shows good agreement with the numerical band. In the 6$^{\rm th}$ band, the FEM shows good agreement with the numerical band, but in the experiment, the periodic plate show does not show much improvement.

 The first band gap periodic plate showed no attenuation in all three cases compared to the solid plate. It was also seen that the band gaps were smaller in the experiment. This is possibly due to the use of a 3$\times$3 finite plate grid. It would be good to examine this using a larger grid (5$\times$5), but the plate size is quite significant to perform experiments in the lab (1.65m $\times$ 1.65m). Although, in theory, the plate should be square, the experimental plate after assembly exhibited a small deviation from the 1:1 aspect ratio. This, along with the nonuniformity of the clamping at some places, leads to the generation of closely spaced frequencies for some modes. Even with these issues, the finite periodic plate does show significant vibration reduction in the predicted band gaps.

\begin{figure}[htpb]
    \centering
    \includegraphics[width=8cm]{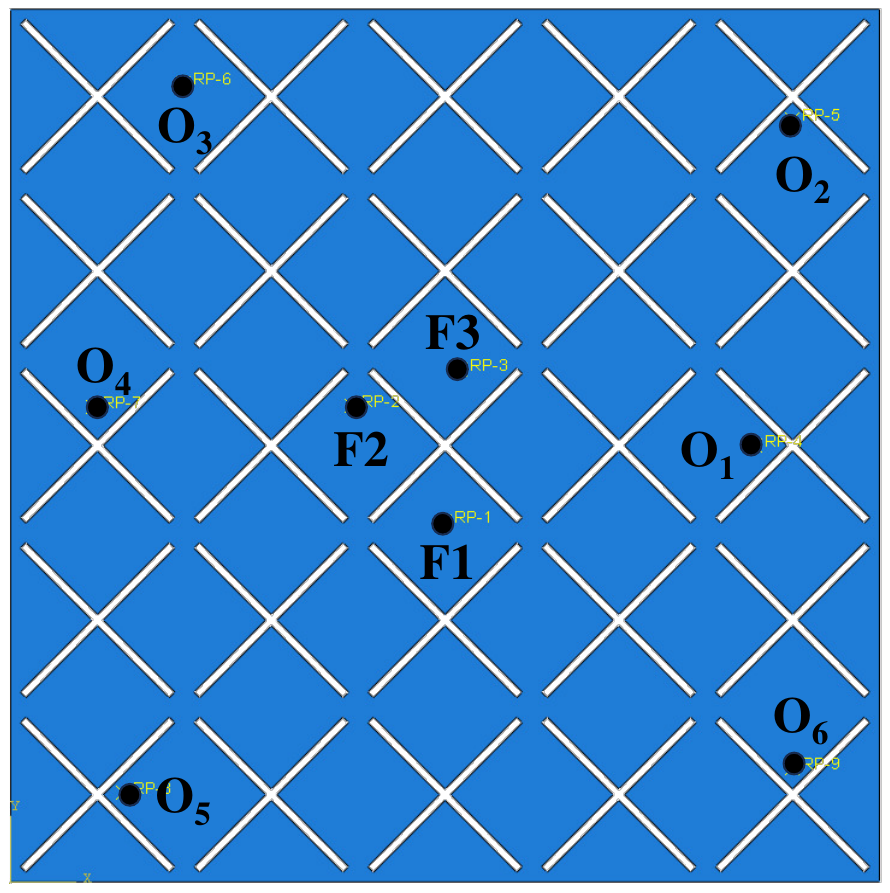}
    \caption{Schematic representation of a 5$\times$5 periodic plate. Note that F$_i,~i=1,2,3$ are the input location and O$_j,~j=1,\cdots,6$ are the locations where the response is measured.}
    \label{fig:5x5}
\end{figure}

\begin{figure}[htpb]
 \centering
  \begin{subfigure}{\textwidth}
        \centering
        \includegraphics[height = 6cm]{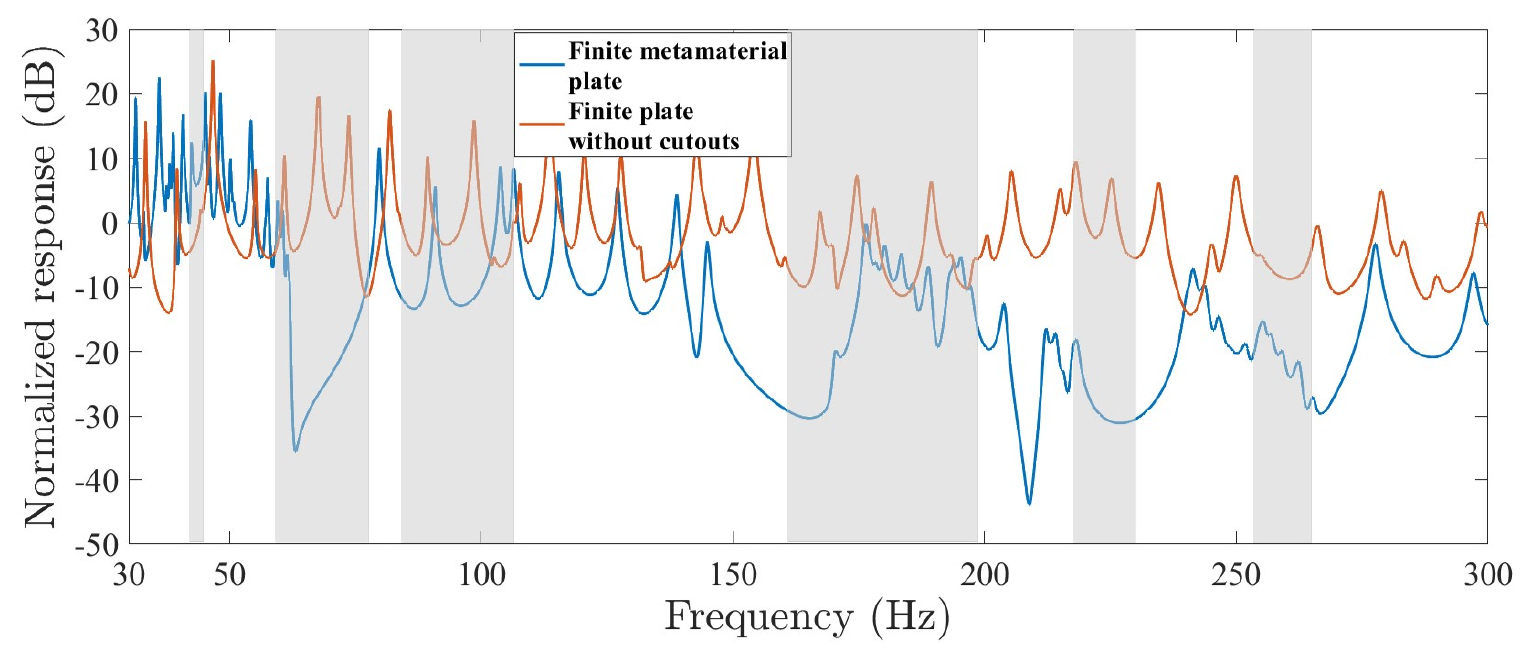}
        \caption{}
        \label{fig:F1_5x5}
  \end{subfigure}
\hfill
    \begin{subfigure}{\textwidth}
        \centering
        \includegraphics[height = 6cm]{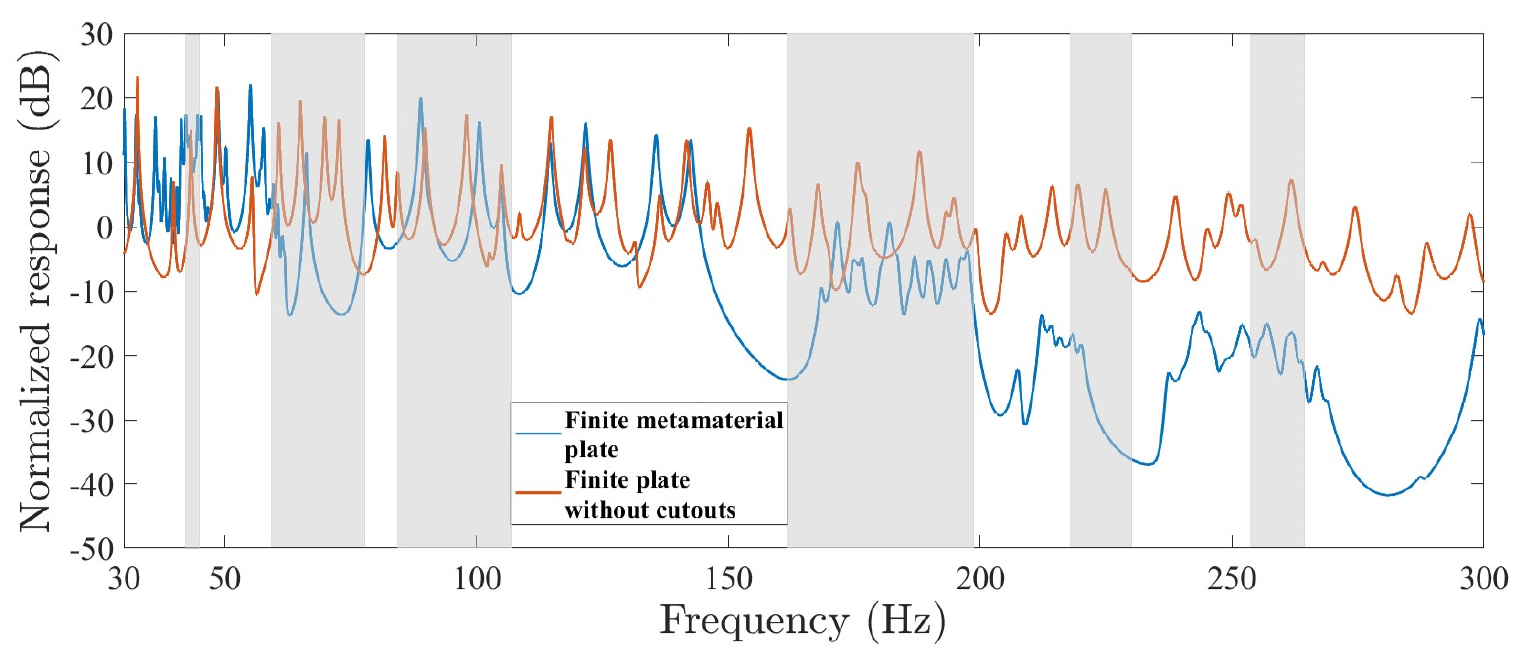}
        \caption{}
        \label{fig:F2_5x5}
    \end{subfigure}
    \hfill
    \begin{subfigure}{\textwidth}
        \centering
        \includegraphics[height = 6cm]{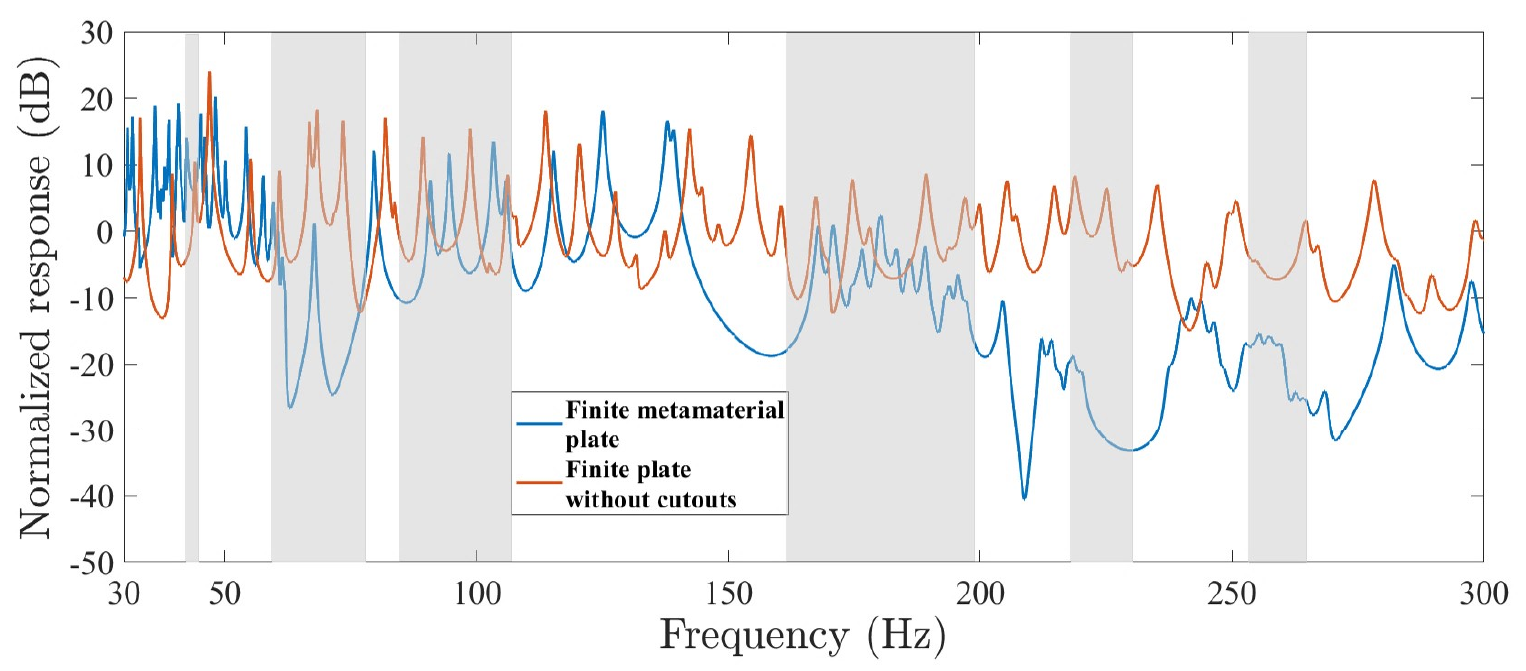}
        \caption{}
        \label{fig:F3_5x5}
    \end{subfigure}
    \caption{5$\times$5 periodic plate normalized response (FEM) for input at (a) F1 (b) F2 (c) F3}
    \label{fig:5x5_Res}
\end{figure}

 A 5$\times$5 periodic plate is examined, similar to the 3$\times$3 plate, using the FE model. \Cref{fig:5x5} shows a 5$\times$5 periodic plate with input and output measuring locations. \Cref{fig:F1_5x5} shows the normalized response for input at location F1. \Cref{fig:F1_5x5} shows that the 2$^{\rm nd}$, 5$^{\rm th}$ and 6$^{\rm th}$ band periodic plates have responses lower than the solid plates. For the 3$^{\rm rd}$ and 4$^{\rm th}$ bands, the periodic plate has a lower response than the solid plate in the frequency range smaller than the band from unit cell studies. \Cref{fig:F2_5x5} shows the response for input at F2. From \Cref{fig:F2_5x5}, it can be seen that for the 2$^{\rm nd}$ band, the periodic plate shows a slightly lower response than the solid plate. Periodic plate show little improvement over solid plate in 3$^{\rm rd}$ band. For the 4$^{\rm th}$ band, the periodic plate shows an overall lower response than the solid plate. In the 5$^{\rm th}$ and 6$^{\rm th}$ bands, periodic plates respond lower than solid plates. The \Cref{fig:F3_5x5} shows the response for input at F3. In the second gap, the periodic plate has a lower response than the solid plate. In the 3$^{\rm rd}$ band, the periodic plate shows no improvement over the solid plate; the same is in the 4$^{\rm th}$ band. Periodic plate shows lower response than solid plate in the 5$^{\rm th}$ and 6$^{\rm th}$ bands. 

 The response of the 5$\times$5 periodic plate shows the same trend as that of the 3$\times$3 periodic plate. However, after 200 Hz, the 5$\times$5 periodic plate shows a considerably reduced response than the 3$\times$3 periodic plate. At lower frequencies, the effect of periodic plate size on the response does not differ much.  The smaller size 3$\times$3 periodic plate shows reasonable attenuation in the frequency range of interest, which makes it practically viable.

\section{Conclusion}
In this paper, the band gap properties of the unit cells with different geometrical shapes for low-frequency regions are studied. The finite element method (FEM) is used to obtain the band gaps. The effect of various geometrical holes on band gaps is discussed. A band gap properties in a finite periodic plate are examined experimentally. From the numerical and experimental study, the following conclusion can be drawn:

\begin{enumerate}
    \item The band gap behavior for different geometries such as ellipse, two diagonal cross ellipses, two diagonal intersecting rectangular slots, intersecting rectangular slots, and offset diagonal rectangular slots have been examined for their band gap properties in the low-frequency region.
    
    \item From the study, it is clear that a high aspect ratio X-type rectangular slots passing through the center are essential to create band gaps in the lower modes. In fact, for the first time, the paper demonstrates the presence of band gaps in the first and second as well as between the third and fourth modes. This has been done keeping the porosity level at around 10\%, which is a significant finding.

    \item A finite plate composed of a 3$\times$3 unit cell arrangement was fabricated and tested for its response to sinusoidal excitation. A FE simulation was also carried out to compare the results obtained from the experiment. Three locations were chosen for the excitation, and the response was averaged over six spread-out locations on the plate. The results demonstrate that the proposed geometry can reduce vibration significantly in the band gap regions.
   
\end{enumerate}

\section*{Data availability}
The data supporting this study's findings are available from the corresponding authors upon reasonable request.

\bibliographystyle{plain}
\bibliography{Ref}

\end{document}